\documentclass{amsart}
\usepackage{graphicx,amssymb}
\newtheorem{thm}{Theorem}[section]
\newtheorem{cor}[thm]{Corollary}

\newtheorem{con}[thm]{Conjecture}
\theoremstyle{definition}
\newtheorem{defn}[thm]{Definition}
\theoremstyle{remark}

\def\beq{\begin{eqnarray}}
\def\eeq{\end{eqnarray}}
\def\bsp{\begin{split}}
\def\esp{\end{split}}

\def\Tr{\mathrm{Tr}}
\def\d{\mathrm{d}}

\def\RC{CSI_R}
\def\KC{CSI_K}
\def\FC{CSI_F}
\newcommand{\FCSI}[1]{CSI_{F,#1}}
\newcommand{\mf}[1]{{\mathfrak #1}}

\newcommand{\mb}[1]{{\mathbb #1}}

\newcommand{\mbold}[1]{\mbox{\boldmath{\ensuremath{#1}}}}

\begin{document}

\title[\textbf{CSI: Halifax}]{\textbf{On Spacetimes with Constant Scalar Invariants}}
\author{\textbf{Alan Coley, Sigbj\o rn Hervik, Nicos Pelavas}}
\address{Department of Mathematics and Statistics,
Dalhousie University, Halifax, Nova Scotia, Canada B3H 3J5}
\email{aac@mathstat.dal.ca, herviks@mathstat.dal.ca,
pelavas@mathstat.dal.ca}
\date{\today}
\maketitle

\begin{abstract}

We study Lorentzian spacetimes for which all scalar invariants
constructed from the Riemann tensor and its covariant derivatives
are constant ($CSI$ spacetimes). We obtain a number of general 
results in arbitrary dimensions. We study and construct
warped product $CSI$ spacetimes and higher-dimensional  Kundt
$CSI$ spacetimes. We show how these spacetimes can
be constructed from locally homogeneous spaces and $VSI$
spacetimes. The results suggest a number of conjectures. In
particular, it is plausible that for $CSI$ spacetimes that are not
locally homogeneous the Weyl type is $II$, $III$, $N$ or $O$, with any
boost weight zero components being constant.
We then consider the four-dimensional $CSI$ spacetimes in more
detail. We show that there are severe constraints on these
spacetimes, and we argue that it is plausible that they are either
locally homogeneous or that  the spacetime necessarily belongs to
the Kundt class of $CSI$ spacetimes, all of which are constructed.
The four-dimensional results lend support to the conjectures in
higher dimensions.

\end{abstract}

\newpage
\section{Introduction}

Let $(\mathcal{M},g)$ denote a differentiable manifold, either of
Riemannian or Lorentzian signature. We
are interested in all spacetimes for which all scalar invariants
constructed from the Riemann tensor and its covariant derivatives
are constant (denoted as $CSI$ spacetimes). In the case of a
Riemannian manifold, if it is $CSI$ then it is locally homogeneous
\cite{PTV}. This theorem is false for Lorentzian manifolds (there
are a number of counterexamples; e.g., the non-homogeneous $VSI$
spacetimes \cite{4DVSI}).

There are a number of $CSI$ spacetimes that arise from homogeneous
spacetimes and simple warped products (see section 3).  Some examples of $CSI$
spacetimes that arise as simple {\it tensor sum} generalizations of
$VSI$ spacetimes with a cosmological constant are given in
\cite{BijPod} (some higher dimensional versions are given in
\cite{obuk}). Simple examples of {\it fibered products} were given
in \cite{Siklos} and \cite{HJS} (see also \cite{Hervik}). It is also straightforward to
construct $CSI$ spacetimes as {\it warped products}. All of these
spacetimes belong to the class $\RC$ described
below.

We begin with some definitions, and a brief discussion of the
four-dimensional case.

\subsection{Notation}

Let us denote by $\mathcal{I}_k$ the set of all scalar
invariants constructed from the curvature tensor and its covariant derivatives up
to order $k$.
\begin{defn}[{$VSI$}$_k$ spacetimes]
$\mathcal{M}$ is called {$VSI$}$_k$ if for any invariant $I\in\mathcal{I}_k$, $I=0$ over $\mathcal{M}$.
\end{defn}
\begin{defn}[{$CSI$}$_k$ spacetimes]
$\mathcal{M}$ is called {$CSI$}$_k$ if for any invariant $I\in\mathcal{I}_k$, $\partial_{\mu}{I}=0$ over $\mathcal{M}$. 
\end{defn}
Moreover, if a spacetime is {$VSI$}$_k$ or {$CSI$}$_k$ for all
$k$, we will simply call the spacetime {$VSI$} or {$CSI$},
respectively. The set of all locally homogeneous spacetimes will be
denoted by $H$. Clearly $VSI \subset CSI$ and $H \subset CSI$.

\begin{defn}[{\sc $\RC$} spacetimes]
Let us denote by $\RC$
all reducible $CSI$ spacetimes that can be built from $VSI$ and
$H$ by (i) warped products (ii) fibered products, and (iii) tensor sums
(defined more precisely later).
\end{defn}

\begin{defn}[{\sc $\FC$} spacetimes]
Let us denote by $\FC$
those spacetimes for which there exists a frame with a null vector
$\ell$ such that all components of the Riemann tensor and its
covariants derivatives in this frame have the property that (i)
all positive boost weight components (with respect to $\ell$) are zero and (ii) all
zero boost weight components are constant.
\end{defn}
Note that $\RC \subset CSI$ and $\FC
\subset CSI$. (There are similar definitions for $\FCSI{k}$
etc. \cite{epsilon}).

\begin{defn}[{$\KC$} spacetimes]
Finally, let us denote by $\KC$, those $CSI$
spacetimes that belong to the (higher-dimensional) Kundt class (defined later);
the so-called Kundt $CSI$ spacetimes.
\end{defn}

In particular, we shall study the relationship between $\RC$,
$\FC$, $\KC$ and especially with
$CSI \backslash H$.  We note that by construction $\RC$ is at
least of Weyl type $II$ (i.e., of type $II$, $III$, $N$ or $O$
\cite{class}), and by definition $\FC$ and
$\KC$ are at least of Weyl type $II$ (more precisely,
at least of Riemann type $II$). In 4D,
$\RC$, $\FC$ and $\KC$ are
closely related, and it is plausible that $CSI \backslash H$ is at least of
Weyl type $II$ (see section 7).

In four dimensions the Weyl classification and the Petrov classification are closely related \cite{class}. However, due to the fact that the Weyl classification contains many more possibilities in dimensions higher than four we will  \emph{restrict the term Petrov classification to four dimensions only}.

\subsection{4D $CSI$}

We are particularly interested in the four-dimensional $CSI$
spacetimes. For a Riemannian manifold every $CSI$ is
homogeneous $(CSI \equiv H)$. This is not true for Lorentzian
manifolds.  However, for every $CSI$ with particular constant
invariants there is a homogeneous spacetime (not necessarily
unique) with precisely the same constant invariants.  This
suggests that $CSI$ can be ``built'' from $H$ and $VSI$ (e.g.,
$\RC$).

In addition, there is a relationship between the $CSI$ conditions
and (i) the rank of the Riemann tensor and its holonomy class  \cite{Hall} (ii)
the existence of curvature collineations \cite{Hall} (iii) the
condition of non-complete backsolvability\footnote{Complete baksolvability (CB) refers to when all of the components of the Riemann tensor can be determined from the CZ set \cite{Carminati} of zeroth order curvature invariants, once all the remaining frame freedom has been used to fix components of the Riemann tensor.} (NCB) \cite{Carminati}, in
addition to (iv) curvature homogeneity\footnote{$\mathcal{M}$ is curvature homogeneous (of order zero) if there exists a frame with respect to which the Riemann tensor has constant components.} and (v) sectional
curvature \cite{Hall}.
The relationship between $CSI$ and curvature homogeneity in $4D$
is studied in \cite{Mp}. In locally homogeneous spacetimes all of the sectional
curvatures (Gaussian curvatures) are constant. With the exception
of certain special plane wave and constant curvature spacetimes
(and all vacuum spacetimes), the sectional curvature uniquely
determines the spacetime metric \cite{Hall}.

It is clear that $CSI$ consists of $\RC$ and,
possibly, some other very special spacetimes. Let us present a
{\em heuristic argument} for this in $4D$. Let us suppose that the
spacetime is $CSI$.  For spacetimes that are completely
backsolvable (CB) there is a special invariant frame such that all
components of the Riemann tensor are constant. With respect to
this invariant frame there exist smooth vector fields $\zeta_i (i
= 1, 2, 3, 4)$ that act transitively on the manifold whose
directional derivatives leave the Riemann tensor invariant (i.e.,
$\zeta_i$ form a Lie algebra of curvature  collineations that span
$\mathcal{M}$).  In general, every curvature collineation is a homothetic
vector, so as a result there is a homothetic group acting
transitively on $\mathcal{M}$. Therefore, this then implies that in general
the $CSI$ spacetime is locally homogeneous.  The exceptions are
those spacetimes that are not CB (NCB \cite{Carminati}) and those
spacetimes for which a curvature collineation is not a homothetic
vector \cite{Hall}, and these spacetimes are related to the $VSI$
spacetimes.

\subsection{Overview}

In this paper we shall study $CSI$ spacetimes. In the
next section we shall begin by summarizing some important results.
In section 3 we construct a subclass of $\RC$
spacetimes that arise as warped products of a homogeneous space and a
$VSI$ spacetime. In section 4 we consider the higher-dimensional
Kundt class. In section 5 examples of $\KC$ spacetimes in
higher dimensions are given. We discuss the results in section 6,
and present a number of conjectures. In section 7 we summarize the
results in 4D in detail.

There are three appendices. In Appendix A, we present all of the
3-dimensional $VSI$ metrics, which is necessary for the explicit
determination of the metrics in the 4-dimensional $\RC$ spacetimes. In
Appendix B,  4-dimensional $CSI$ spacetimes are given explicitly,
and the relationship between $\RC$,
$\FC$, and $\KC$ in 4D is discussed.
Finally, in Appendix C, we write down the metric for higher
dimensional $VSI$ spacetimes in a canonical form.

\section{Spacetimes with constant scalar invariants}

\subsection{Riemannian case}
Let us first consider the Riemannian case where a great deal about these spacetimes is known. The reason the Riemannian case is easier to deal with is because the orthogonal group $O(d)$ is compact, and hence, the orbits of its group action are also compact.

\begin{thm}
If $\mathcal{M}$ is a Riemannian spacetime, then:
\begin{enumerate}
\item{} {$VSI$}$_0$ $\Rightarrow$ {$VSI$}.
\item{} There exists a $k\in\mathbb{N}$ such that {$CSI$}$_k$ $\Rightarrow$ {$CSI$}.
\end{enumerate}
\end{thm}
\begin{proof}
(1): See proof below. \quad (2): See \cite{PTV}
\end{proof}
\begin{thm}[Riemannian {$VSI$}]
A Riemannian space is {$VSI$} if and only if it is flat.
\end{thm}
\begin{proof}
Consider the invariant $R_{ABCD}R^{ABCD}$. Using an orthonormal frame, this invariant is a sum of squares; hence, $R_{ABCD}R^{ABCD}=0$ $\Leftrightarrow$ $R_{ABCD}=0$.
\end{proof}
\begin{thm}[Riemannian {$CSI$}]
A Riemannian space is {$CSI$} if and only if it is locally homogeneous. Moreover, the set of curvature invariants, $\mathcal{I}$, uniquely determines the metric up to  isometries.
\end{thm}
\begin{proof}
See Pr\"ufer-Tricerri-Vanhecke \cite{PTV}.
\end{proof}

\begin{thm}
[Singer]  If $\mathcal{M}$ is curvature homogeneous of order $k \geq
\frac{1}{2} n(n-1)$, then $\mathcal{M}$ is locally homogeneous.
\end{thm}
\begin{proof}
See Singer \cite{Singer}.
\end{proof}

\subsection{Lorentzian case}
The Lorentzian case is more difficult to deal with and only
partial results are known. However, in the {$VSI$} case, it was
shown in \cite{Higher} that {$VSI$}$_2$ implies {$VSI$}. A
similar result in the {$CSI$} case is not known;
however, in this case we believe that an $n$ exists such that
{$CSI$}$_n$ implies {$CSI$}.

In the Riemannian case $\mathcal{M}$ is locally homogeneous if and only if $\mathcal{M}$ is
curvature homogeneous to all orders (Theorem 2.4). In the
Lorentzian case, we have that
\begin{thm}[Podesta and Spiro]
There is an integer $K_{p,q}$ such that if $(M,g)$ is a complete,
simply connected pseudo Riemannian manifold of type $(p,q)$ which
is $K_{p,q}$-curvature homogeneous, then $\mathcal{M}$ is locally
homogeneous.
\end{thm}
\begin{proof}
See Podesta and Spiro \cite{PS}.
\end{proof}

We argued above that in 4D Lorentzian manifolds for every $CSI$ with particular constant
invariants there is a locally homogeneous spacetime with 
the same constant invariants.  It is plausible that this is
true in higher dimensions.

\begin{con}
Assume that a  Lorentzian spacetime $\mathcal{M}$ is a {$CSI$} spacetime with
curvature invariants $\mathcal{I}$. Then there exists a locally homogeneous
space $\widetilde{\mathcal{M}}$ with curvature invariants
$\widetilde{\mathcal{I}}=\mathcal{I}$. 
\end{con}

\subsection{Boost-weight decomposition}
Consider an arbitrary covariant tensor $T$ and a null frame $\left\{\ell_{A},n_{A}, m^{\hat\mu}_{A}\right\}$; i.e.,
\[ \ell_{A}n^{A}=1,~~m^{\hat\mu}_{~A}m^{\hat\nu A}=\delta^{\hat\mu\hat\nu}, ~~ \ell_{A}\ell^{A}=n_{A}n^{A}=\ell_{A}m^{\hat\mu A}=
n_{A}m^{\hat\mu A}=0.\]
Consider now a boost in the $\ell^{A}-n^{A}$-plane:
\beq
\left\{\tilde{\ell}_{A},\tilde{n}_{A}, \tilde{m}^{\hat\mu}_{~A}\right\}= \left\{e^{\lambda}\ell_{A},e^{-\lambda}n_{A}, m^{\hat\mu}_{~A}\right\}.
\eeq
We can consider the vector-space decomposition of the tensor $T$ in terms of the boost weight with respect to the above boost (see \cite{Milson}):
\beq
T=\sum_b(T)_b,
\eeq
where $(T)_b$ denotes the projection of the tensor $T$ onto the vector space of boost-weight $b$. The components of the tensor $(T)_b$ with respect to the frame will transform according to:
\beq
(T)_{b~AB...}=e^{-b\lambda}(T)_{b~\tilde{A}\tilde{B}...}.
\eeq
Furthermore, we note that
\beq
\left(T\otimes S\right)_b=\sum_{b'+b''=b}(T)_{b'}\otimes(S)_{b''}.
\eeq
Moreover, the connection  $\nabla$ can similarly be decomposed according to whether it raises, lowers or preserves the boost weight:
\beq
\nabla=(\nabla)_{-1}+(\nabla)_0+(\nabla)_1,
\eeq
where $(\nabla)_b$ is defined by
\[ (\nabla)_{b{\bf X}}\equiv \nabla_{({\bf X})_b},\]
for all vectors ${\bf X}$. 
We note that for the metric, $g=(g)_0$; hence, raising or lowering tensor indices preserves the boost weight.

An invariant of a tensor, $T$, is necessarily $SO(1,n)$-invariant; in particular, for any full contraction of $T$:
\beq
\mathrm{Cont}[T]=\mathrm{Cont}[(T)_0].
\eeq
This property can now be utilized to construct spaces with constant curvature invariants. In the {$VSI$} case, there exists a null-frame such that the Riemann tensor, $R$, has the property:
\beq
(R)_b=0,\quad b\geq 0.
\eeq
This property alone implies that $\mathcal{I}_0=0$, regardless of $(R)_b$ ($b<0$). 
If the curvature tensor is of the form:
\beq
(R)_b=0,\quad b> 0, \quad \text{and}~(R)_{b~ABCD,E}=0,\quad b= 0,
\eeq
in a frame for which $g_{AB,E}=0$, it is necessarily a {$CSI$}$_0$ spacetime. In general, further restrictions must be put on
$\nabla$ in order to make it {$CSI$}. We note that spacetimes can can be classified according 
to their boost-weight components $(R)_b$. In particular, a classification using the Weyl 
tensor $C$, which generalises the Petrov classification in 4D, has been employed based on 
the components $(C)_b$ \cite{class,Milson}.

\subsection{Kundt metrics}
\label{sect:Kundt} In the 4D $VSI$ spacetimes, all of the positive
boost weight spin coefficients (e.g., $\kappa, \rho$ and $\sigma$)
are zero, and hence it follows that $\ell$ is geodesic,
non-expanding, shear-free and non-twisting and the spacetime is
Kundt $K$ \cite{exsol}. This is also true for $CSI$ spacetimes,
in the sense that if the $CSI$ spacetime is not locally
homogeneous, then in general it belongs to $K$. That is, it is plausible
that in 4D, $CSI \equiv H \cup \KC$.

In higher dimensions it was also shown that $\ell$ is geodesic,
non-expanding, shear-free and non-twisting (i.e., $L_{ij}=0$) in
$VSI$ spacetimes \cite{Higher}. In locally homogeneous spacetimes, in general 
there exists a null frame in which the
$L_{ij}$ are constants. We therefore anticipate that in $CSI$
spacetimes that are not locally homogeneous, $L_{ij}=0$. For
higher dimensional $CSI$ spacetimes with $L_{ij}=0$, the Ricci and
Bianchi identities appear to be identically satisfied
\cite{bianchi}. A higher dimensional spacetime which admits a null
vector $\ell$ which is geodesic, non-expanding, shear-free and
non-twisting, will be denoted as a higher-dimensional Kundt
spacetime.

Let us assume that the spacetime admits a geodesic, non-twisting, non-expanding,
shear-free null vector ${\ell}$. It can be then shown that the existence of a twist-free and geodesic null vector implies that there exists a local coordinate system $(u,v,x^i)$ such that
\beq
\d s^2=2\d u\left(\d
v+H\d u+W_{ i}\d x^i\right)+\tilde{g}_{ij}(v,u,x^k)\d x^i\d x^j.
\eeq
In essence, the twist-free condition implies that the vector field is 'surface forming'
so that there exists locally an exact null one-form $\d u$.
Using a coordinate transformation and a null-rotation we can now bring the metric into
the above form. In this coordinate system, $\ell=\frac{\partial}{\partial v}$.
In addition, requiring that this vector field is non-expanding and shear-free implies
that $\tilde{g}_{ij,v}=0$.

We will therefore consider metrics of the from
\beq \d s^2=2\d u\left(\d
v+H\d u+W_{ i}\d x^i\right)+\tilde{g}_{ij}(u,x^k)\d x^i\d x^j,
\label{Kundt}
\eeq
where  $H=H(v,u,x^k)$ and
$W_{i}=W_{i}(v,u,x^k)$. The metric (\ref{Kundt})
possesses a null vector field $\ell$ obeying \beq
\ell^{A}\ell_{B;A}=\ell^{A}_{~;A}=\ell^{A;B}\ell_{(A;B)}=\ell^{A;B}\ell_{[A;B]}=0;
\eeq i.e., it is geodesic, non-expanding, shear-free and
non-twisting. We will refer to the metrics (\ref{Kundt}) as
higher-dimensional \emph{Kundt metrics} (or simply Kundt metrics),
since they generalise the 4-dimensional Kundt metrics.

\subsection{Lorentzian {$CSI$}} All examples to the authors knowledge
of Lorentzian {$CSI$} spacetimes are of the following two forms:
\begin{enumerate}
 \item{}Homogeneous spaces
 \item{}A subclass of the Kundt spacetimes
\end{enumerate}
For the homogeneous spaces, their decomposition can be of the most general type,
\[ \nabla^{(k)}R= \sum_{b=-(2+k)}^{2+k}\left(\nabla^{(k)}R\right)_b, \]
since a frame can be chosen such that all components are constants. However, for all known examples of non-homogeneous {$CSI$} metrics, a frame can be found such that we have the decomposition
\[  \nabla^{(k)}R= \sum_{b=-(2+k)}^{0}\left(\nabla^{(k)}R\right)_b, \]
with constant boost-weight zero components. 

Since a homogeneous space is automatically a {$CSI$} space, we
will in this paper consider metrics which are not necessarily
homogeneous. Since all known examples of non-homogeneous metrics
are of Kundt form, we will henceforth only consider Kundt metrics.

\section{Warped product {$CSI$}}

It is known that  $\RC$ spacetimes can be
constructed in all dimensions via warped products. Let us first
determine necessary and sufficient conditions so that the warped
product of a homogeneous space and a $VSI$ is $CSI$.

We consider the warped product metric $g_{AB}=\stackrel{1}{g}_{ab}
\oplus\; e^{2\tau}\hspace{-0.1cm}\stackrel{2}{g}_{\alpha\beta}$
where $\stackrel{1}{g}$ is a Riemannian homogenous space and
$\stackrel{2}{g}$ is a $k$-dimensional {$VSI$} Lorentzian
manifold, the warping function $\tau$ only depends on points of
the homogenous space.  Using the left-invariant
frame of the homogeneous space\footnote{Hatted indices are the
preferred null-frame indices.},

\begin{equation}
{\sf e}_{\hat a}=\left\{{\bf m}_{\hat k},\ldots,{\bf m}_{\widehat{N-1}}\right\},
\end{equation}
we complete it to a basis of the warped product by appending the null frame,

\begin{equation}
{\sf e}_{\hat\alpha}=\left\{ {\bf \ell}=e^{-\tau}{\bf \ell'},{\bf n}=e^{-\tau}{\bf n'},{\bf m}_{\hat 2}=e^{-\tau}{\bf
m'}_{\hat 2},\ldots,{\bf m}_{\widehat{k-1}}=e^{-\tau}{\bf m'}_{\widehat{k-1}} \right\},
\end{equation}
where the primed vectors represent the canonical {$VSI$} frame in which ${\bf \ell'}$ (and hence $\ell$) is the aligned
geodesic null congruence that is expansion, shear, and twist free. The non-vanishing inner products are such that

\begin{equation}
g_{AB}=2\ell_{(A}n_{B)}+\delta_{\hat\mu\hat\nu}m^{\hat\mu}_{~A}m^{\hat\nu}_{~B}+\delta_{\hat m\hat n}m^{\hat m}_{~A}m^{\hat n}_{~B}.
\end{equation}

Defining $T_{\hat m\hat p}=\tau_{,\hat m ; \hat p}+\tau_{,\hat m}\tau_{,\hat p}$, we shall show that if $\tau_{,m}\tau^{,m}$ and $T_{\hat m
\hat p}$ are constant then this implies {$CSI$}.  These two conditions imply that $\tau^{,\hat m}$ is an eigenvector of $T_{\hat m
\hat p}$ with constant eigenvalue $\tau_{,m}\tau^{,m}$, and thus $\tau_{,\hat b}$ must be constant.

Following \cite{Higher}, we perform a decomposition of the Riemann
tensor for the warped product to obtain

\begin{eqnarray}
R_{ABCD} = 4R_{\hat 0\hat 1\hat 0\hat 1}n_{\{A}\ell_{B}n_{C}\ell_{D\}}+ 8R_{\hat 0\hat \mu\hat 1\hat\lambda}n_{\{A}m^{\hat \mu}_{~
B}\ell_{C}m^{\hat \lambda}_{~ D\}}+
8R_{\hat 0\hat a\hat 1\hat b}n_{\{A}m^{\hat a}_{~B}\ell_{C}m^{\hat b}_{~D\}} \nonumber \\
+R_{\hat \mu\hat \nu\hat \lambda\hat \sigma}m^{\hat \mu}_{~\{A}m^{\hat\nu}_{~B}m^{\hat\lambda}_{~ C}m^{\hat\sigma}_{~D\}}
+4R_{\hat a\hat\mu\hat b\hat\lambda}m^{\hat a}_{~\{A}m^{\hat\mu}_{~ B}m^{\hat b}_{~ C}m^{\hat\lambda}_{~ D\}} \label{riem}\\
+R_{\hat a\hat b\hat c\hat d}m^{\hat a}_{~ \{A}m^{\hat b}_{~ B}m^{\hat c}_{~ C}m^{\hat d}_{ D\}}+\text{negative boost weight terms}\cdots  \nonumber
\end{eqnarray}
Evidently the Riemann tensor is of boost order zero with boost weight zero components:

\begin{eqnarray}
R_{\hat 0\hat 1\hat 0\hat 1} &=& \tau_{,m}\tau^{,m}       \label{rbw0.1} \\
R_{\hat 0\hat \mu\hat 1\hat \lambda} &=& -\tau_{,m}\tau^{,m}\stackrel{2}{\delta}_{\hat \mu\hat\lambda}  \label{rbw0.2}  \\
R_{\hat 0\hat a\hat 1\hat b} &=&  -T_{\hat a\hat b}                       \label{rbw0.3} \\
R_{\hat\mu\hat\nu\hat\lambda\hat\sigma} &=& \tau_{,m}\tau^{,m}
\left[\stackrel{2}{\delta}_{\hat\nu\hat\lambda}\stackrel{2}{\delta}_{\hat\mu\hat\sigma} -
\stackrel{2}{\delta}_{\hat\mu\hat\lambda}\stackrel{2}{\delta}_{\hat\nu\hat\sigma} \right]  \label{rbw0.4} \\
R_{\hat a\hat\mu\hat b\hat\lambda} &=& - T_{\hat a\hat b}\stackrel{2}{\delta}_{\hat\mu\hat\lambda} \label{rbw0.5} \\
R_{\hat a\hat b\hat c\hat d} &=& \stackrel{1}{R}_{\hat a\hat b\hat c\hat d}.  \label{rbw0.6}
\end{eqnarray}
The negative boost weight components, arising from the {$VSI$} spacetime, are given by
\begin{eqnarray}
R_{\hat 1\hat 0\hat 1\hat \mu} &=& e^{-2\tau}\stackrel{2}{R}_{\hat 1\hat 0\hat 1\hat \mu}  \label{rbw-1.4}  \\
R_{\hat 1\hat\mu\hat\nu\hat\lambda} &=& e^{-2\tau}\stackrel{2}{R}_{\hat 1\hat\mu\hat\nu\hat\lambda}  \label{rbw-1.5}  \\
R_{\hat 1\hat\mu\hat 1\hat\nu} &=& e^{-2\tau}\stackrel{2}{R}_{\hat 1\hat\mu\hat 1\hat\nu}.  \label{rbw-2.6}
\end{eqnarray}
The condition that $\tau_{,\hat b}$ is constant (and hence $\tau_{,m}\tau^{,m}$ and 
$T_{\hat m\hat p}$ are constant) implies that
the boost weight zero components of the Riemann tensor are constant; the absence of positive boost weight components
then gives {$CSI$}$_{0}$.

The covariant derivative of the null frame has the form,
\begin{eqnarray}
\ell_{K;E} &=& -\tau_{,\hat b}m^{\hat b}_{~ K}\ell_{E}+\gamma_{\hat\lambda\hat 0\hat 1}m^{\hat\lambda}_{~ K}\ell_{E}
+\gamma_{\hat 1\hat 0\hat \lambda}\ell_{K}m^{\hat\lambda}_{~E} + \cdots \label{dl} \\ 
n_{K;E} &=& -\tau_{,\hat b}m^{\hat b}_{~ K}n_{E}-\gamma_{\hat 1\hat 0\hat \lambda}n_{K}m^{\hat \lambda}_{~ E} + \cdots \label{dn}\\
m_{\hat \lambda K;E} &=& -\tau_{,\hat b}m^{\hat b}_{~K}m_{\hat \lambda E}+\gamma_{\hat \mu\hat \lambda\hat \nu}m^{\hat\mu}_{~
K}m^{\hat \nu}_{~E} - \gamma_{\hat \lambda\hat 0\hat 1}n_{K}\ell_{E}+ \cdots \label{dmv}\\
m_{\hat a K;E} &=& \tau_{,\hat a}\ell_{K}n_{E}+\tau_{,\hat a}n_{K}\ell_{E} +\tau_{,\hat a}\delta_{\hat \mu\hat\lambda}m^{\hat\mu}_{~
K}m^{\hat\lambda}_{~ E}+\stackrel{1}{\gamma}_{\hat b\hat a\hat c}m^{\hat b}_{~K}m^{\hat c}_{~ E} \label{dma}
\end{eqnarray}
where $\cdots$ denotes terms of lower order boost weight with
respect to previous terms in an expression, and the unspecified
$\gamma$'s correspond to {$VSI$} Ricci-rotation coefficients
\cite{Higher} [for example, $\gamma_{\hat 1\hat 0\hat
\lambda}=e^{-\tau}L_{\hat 1\hat\lambda}$], except for the
$\stackrel{1}{\gamma}$ which are the rotation coefficients
associated with the homogeneous space. Equations
(\ref{dl})-(\ref{dma}) have leading order boost weight terms of
$+1,~-1,~0$ and $0$, respectively, which is the same boost weight
as the corresponding null frame vector before taking a covariant
derivative.  It follows that the covariant derivative of the
$\{\cdot\}$ quantities appearing in (\ref{riem}) will produce
terms of equal or lesser boost weight.  Furthermore, in
\cite{Higher} it was proven that the negative boost weight
curvature components appearing in (\ref{riem}) will remain
negative upon covariant differentiation.  Therefore, if the boost
weight zero components of the Riemann tensor and its derivatives
are constant then any number of covariant derivatives of the
Riemann tensor will be of boost order zero, which implies that the
warped product will be {$CSI$}.

Supposing that the boost weight zero components of $\nabla^{(n)}R$ are constant, then for $\nabla^{(n+1)}R$ the boost
weight zero components will have contributions from $\tau_{,\hat a}$, which is constant, and the non-constant $\gamma$'s
arising from the {$VSI$} spacetime, however these boost weight zero {$VSI$} 
$\gamma$'s do not occur as a result of
Theorem 5.1.  Therefore we have that if $\tau_{,\hat b}$ is constant, then the warped product is {$CSI$}.

For reference, we  include the covariant derivative of the Riemann tensor.  By using the Ricci identity we have,
\begin{equation}
\tau^{,\hat a}\stackrel{1}{R}_{\hat a\hat b\hat c\hat d}+2T_{\hat b[\hat d}\tau_{,\hat c]}=T_{\hat b\hat c;\hat d}-T_{\hat b\hat d;\hat c},   \label{riciden}
\end{equation}
which can be used to express the non-vanishing boost weight zero components of the covariant derivative of the Riemann
tensor as
\begin{eqnarray}
R_{\hat 1\hat b\hat c\hat d;\hat 0} &=& T_{\hat b\hat c;\hat d}-T_{\hat b\hat d;\hat c}=R_{\hat 0\hat b\hat c\hat d;\hat 1} \label{drbw0.1} \\
R_{\hat \mu\hat b\hat c\hat d;\hat \lambda} &=& \left[T_{\hat b\hat c;\hat d}-T_{\hat b\hat d;\hat c}\right]\delta_{\hat\mu\hat\lambda} \label{drbw0.2} \\
R_{\hat 0\hat b\hat 1\hat c;\hat e} &=& -T_{\hat b\hat c;\hat e}   \label{drbw0.3} \\
R_{\hat b\hat \mu\hat c\hat \lambda;\hat e} &=& -T_{\hat b\hat c;\hat e}\delta_{\hat \mu\hat \lambda}.   \label{drbw0.4}
\end{eqnarray}
As expected all components in (\ref{drbw0.1})-(\ref{drbw0.4}) are constant (as a result of requiring $\tau_{,\hat b}$ to
be constant).

We shall now show the converse; that is, if the Ricci invariants are constant then $\tau_{,\hat b}$ is constant.
In local coordinates, the Ricci tensor and Ricci scalar are
\begin{eqnarray}
R_{\mu}^{~\rho} &=& e^{-2\tau}\stackrel{2}{R}_{\mu}^{~\rho}
-[T+(k-1)\tau_{,m}\tau^{,m}]\delta_{\mu}^{\rho} \label{ric2} \\
R_{m}^{~ p} &=& \stackrel{1}{R}^{~ p}_{m}-kT_{m}^{~ p}  \label{ric1} \\
R &=&\stackrel{1}{R}-k[2T+(k-1)\tau_{,m}\tau^{,m}],\label{ricsc}
\end{eqnarray}
where in (\ref{ricsc}) we have used that $\stackrel{2}{R}=0$ and set $T=\Tr(T_{m}^{~p})$. Choosing new coordinates
$\tilde{x}^{1}=\tau(x^{m}), \tilde{x}^{2}=x^{2},\ldots$ so 
that $\tau_{,m}\tau^{,m}=\tilde{g}^{11}$, then

\begin{equation}
\tilde{T}_{mn}= -\tilde{\Gamma}^{1}_{mn}+\delta^{1}_{m}\delta^{1}_{n}.
\end{equation}

Now, with normal coordinates along $\tilde{x}^{1}$ we have $\tilde{g}^{11}=1$.  Therefore from (\ref{ricsc}), since
$\stackrel{1}{R}$ is constant we have that $\tilde{T}$, and hence $T$, is constant.  In this coordinate system the
constraint $\tilde{T}$ constant implies that $\tilde{g}^{mn}\tilde{g}_{mn,1}$ is constant.  Constant Ricci scalar
implies that $T$ and $\tau_{,m}\tau^{,m}$ are constant. Further conditions are then provided by the constancy of higher
degree Ricci invariants.

From (\ref{ric2}) and (\ref{ric1}),  the Ricci tensor of a warped product is $R_{M}^{\ \ N}=R_{m}^{\ \ n} \oplus
R_{\mu}^{\ \ \nu}$, thus a degree $d$ Ricci invariant always has the form

\begin{equation}
R_{N_{1}}^{\ \ N_{2}}R_{N_{2}}^{\ \ N_{3}}\cdots R_{N_{d}}^{\ \ N_{1}}=R_{n_{1}}^{\ \ n_{2}}R_{n_{2}}^{\ \ n_{3}}\cdots
R_{n_{d}}^{\ \ n_{1}} + R_{\nu_{1}}^{\ \ \nu_{2}}R_{\nu_{2}}^{\ \ \nu_{3}}\cdots R_{\nu_{d}}^{\ \ \nu_{1}}.
\label{ricinv}
\end{equation}
The second term of (\ref{ricinv}) is constant since $T$ and $\tau_{,m}\tau^{,m}$ are constant; hence invariants
constructed from (\ref{ric1}) must be constant. Assuming the determinant of $R_{m}^{ ~p}$ or $\stackrel{1}R^{~
p}_{m}$ is nonzero, then there exists a frame in which $R_{m}^{~ p}$ and $\stackrel{1}R^{~ p}_{m}$ can
simultaneously be expressed in block diagonal form with constant eigenvalues \cite{Petrov}. Therefore, (\ref{ric1}) implies that in
this frame $T_{m}^{~ p}$ must also have constant components and thus constant invariants. 
In this case, the constancy of the Ricci
invariants completely determines the conditions on $T_{m}^{~ p}$ and specifies a frame in which the
components of each term in (\ref{ric1}) are constant. The frame in which $\stackrel{1}R^{~ p}_{m}$ and $T_{m}^{~ p}$ are
constant is determined by the left-invariant 1-forms.  As before, the conditions that $T_{\hat a\hat b}$ and
$\tau_{,m}\tau^{,m}$ are constant imply $\tau_{,\hat a}$ is constant, and consequently 
constant Ricci invariants implies {$CSI$}.

Many examples of {$CSI$} spacetimes can be found among the cases where the homogeneous space is a solvable Lie group\footnote{Recall that a Lie group can be equipped with a left-invariant metric (i.e., invariant under the left action) in the standard way, see e.g. \cite{Milnor}.}:
\begin{thm}
Assume that $\mathcal{M}_H$ is a solvable Lie group equipped with a left-invariant metric. Assume also that $\mathcal{M}_H$ is connected and simply connected (i.e., $H^1(\mathcal{M}_H)=0$). Then there exists a non-constant $\tau$ such that the warped product of $\mathcal{M}_H$ and a $VSI$, as constructed above, is a {$CSI$}.
\end{thm}
\begin{proof}
We identify $\mathcal{M}_H=G$, where $G$ is a solvable Lie group, with the Lie algebra $\mf{g}=T_{e}G$. Since $\mf{g}$ is solvable the set $[\mf{g},\mf{g}]$ will be a proper vector subspace of $\mf{g}$. Thus for a solvable Lie algebra, there exists a non-zero ${\bf X}\in\mf{g}\cap[\mf{g},\mf{g}]^{\perp}$ where $[\mf{g},\mf{g}]^{\perp}$ is the complement of the derived algebra $[\mf{g},\mf{g}]$. This further implies that there is a left-invariant one-form ${\mbold\omega}$ such that $\d {\mbold\omega}=0$. Since $H^1(\mathcal{M}_H)=0$, there exists a non-constant function $\phi$ such that ${\mbold\omega}=\d \phi$. We can now choose $\tau=p\phi$ for a constant $p$.
\end{proof}

In 2-dimensions flat space is the only {$VSI$} metric, and in
the appendix we list all of the 3-dimensional {$VSI$} metrics.
Examples of {$CSI$} warped products where the homogeneous space
is either $\mathbb{E}^n$ or $\mathbb{H}^n$ yield a linear function
for $\tau$ (since both of these can be considered as
solvmanifolds); however, for $\mathbb{S}^n$, since the coordinates
are cyclic, we find that $\tau$ is a constant. Another example, in
5-dimensions, is where the homogeneous space is a Bianchi type VIII Lie group ($\cong SL(2,\mathbb{R})$)
warped with a 2-dimensional {$VSI$}, with coordinates $u,v,x,y,z$
the metric is
\beq
\d s^2 &=& e^{2\tau}\left[2\d u(\d v+H\d u)\right]\nonumber \\
&& +\frac{a^2}{y^2}(\cos z\d x+\sin z\d y)^2+\frac{b^2}{y^2}(\cos z\d y-\sin z\d x)^2+c^2\left(\d z+\frac{\d x}{y}\right)^2.
\eeq
In this case $H(u,v)$ must be linear in $v$ to give {$CSI$}, and it turns out that for the warped product to be {$CSI$} we have to consider two separate cases:
\begin{enumerate}
\item{} The maximal isometry group of the homogeneous space is 3-dimensional ($a\neq b$): $\tau$ is a constant.
\item{} The maximal isometry group of the homogeneous space is 4-dimensional ($a=b$): $\tau$ can be non-constant, $\tau=\alpha \ln y+\beta z$.
\end{enumerate}
The reason why we have to distinguish these to cases is that in the first case, the isometry group is the semi-simple group, $SL(2,\mathbb R)$. However, in the
second case, the isometry group allows for a transitive group
which is solvable (Bianchi type III).

\section{Kundt {$CSI$} metrics}

Let us next consider the higher-dimensional Kundt metrics (\ref{Kundt})
in the form
\beq \d
s^2=2\d u\left(\d v+H\d u+W_{\hat i}{\bf m}^{\hat
i}\right)+\delta_{\hat i\hat j}{\bf m}^{\hat i}{\bf m}^{\hat j}.
\label{Kundtmetric}\eeq 
It is convenient to introduce the null frame
\beq
\ell&=& \d u, \\
{\bf n}&=& \d v+H\d u+W_{\hat i}{{\bf m}}^{\hat i}, \\
{\bf m}^{\hat i}&=& {\sf e}^{\hat i}_{~j}(u; x^{k})\d x^j,\\
(\eta_{AB}) &=&\begin{bmatrix}
0 & 1 & 0 \\
1 & 0 & 0 \\
0 & 0 & \delta_{\hat i\hat j}
\end{bmatrix}
\eeq
such that
\beq
\d S^2_H\equiv\delta_{\hat i\hat j}{\bf m}^{\hat i}{\bf m}^{\hat j}=\tilde{g}_{ij}\d x^i\d x^j,\quad \tilde{g}_{ij,v}=0.
\eeq

There is a class of coordinate transformations that preserve the form of the metric (\ref{Kundtmetric}). In particular, we can define new variables,
\beq
(v',u',x'^i)=(v,u,f^i(u;x^k)).
\eeq
For our purposes we can always use this coordinate transformation to simplify the spatial metric $\tilde{g}_{ij}$:
\begin{thm}
Consider the metric (\ref{Kundt}) and assume that there exists a null frame $\left\{\ell_A,n_A,m^{\hat{i}}_{~A}\right\}$ such that all scalars $R_{\hat{i}\hat{j}\hat{k}\hat{l}}\equiv R_{ABDC}m^{~A}_{\hat{i}}m^{~B}_{\hat{j}}m^{~C}_{\hat{k}}m^{~D}_{\hat{l}}$ and $R_{\hat{i}\cdots;\hat{j}_1\cdots\hat{j}_n}\equiv R_{A\cdots ;C_1\cdots C_n }m^{~A}_{\hat{i}}\cdots m^{~C_1}_{\hat{j}_1}\cdots m^{~C_n}_{\hat{j}_n}$ are constants. Then there exists (locally) a coordinate transformation $(v',u',x'^i)=(v,u,f^i(u;x^k))$ such that
\beq
\tilde{g}_{ij}\equiv \tilde{g}'_{kl}\frac{\partial f^k}{\partial x^i}\frac{\partial f^l}{\partial x^j}, \quad \tilde{g}'_{ij,u'}=0.
\eeq
Moreover, $\d S_H^2=\tilde{g}'_{ij}\d x'^i\d x'^j$ is a locally homogeneous space.
\label{thm:Kundt}
\end{thm}
\begin{proof}
By calculating the curvature tensor of the metric (\ref{Kundt}), and its covariant derivatives, we note that
\beq
R_{\hat{i}\hat{j}\hat{k}\hat{l}}=\widetilde{R}_{\hat{i}\hat{j}\hat{k}\hat{l}}, \quad R_{\hat{i}\cdots;\hat{j}_1\cdots\hat{j}_n}=\widetilde{R}_{\hat{i}\cdots;\hat{j_1}\cdots\hat{j}_n},
\eeq
where the tilde refers to curvature tensors with respect to the spatial metric $\tilde{g}_{ij}(u,x^k)$. Since this metric is Riemannian with all its components constant we can use the results of \cite{PTV} which state that the metrics $\tilde{g}_{ij}(u,x^k)$, for different $u$, are equivalent up to isometries. Thus, given $u_0$, there exists for every $u$ sufficiently close to $u_0$ an isometry $\psi_u(x^k)$ such that $(\psi_u^*\tilde{g})(u)=\tilde{g}(u_0)$. Since, we assume that $\tilde{g}$ is sufficiently smooth in a neighbourhood of $u_0$, we can find a sufficently smooth $\psi_u$ in $u$. The map $\psi_u(x^k)$ provides us with the functions $f^i$ in the theorem by composition  with the coordinate charts. Hence, we have $\tilde{g}'=\tilde{g}\big|_{u=u_0}$. Moreover, $\tilde{g}'$ is a locally homogeneous space \cite{PTV}.
\end{proof}

Henceforth, we shall assume that we consider solutions in the set $\FC  \bigcap \KC$.
Consequently, from Theorem 4.1, there is no loss of generality in assuming that the metric 
(\ref{Kundt}) has $\tilde{g}_{ij,u}=0$. 
The remaining coordinate freedom preserving the Kundt form is then:
\begin{enumerate}
\item{} $(v',u',x'^i)=(v,u,f^i({x}^k))$ and $J^i_{~j}\equiv \frac{\partial f^i}{\partial x^j}$.
\beq
H'= H, \quad W'_i=W_j\left(J^{-1}\right)^j_{~i}, \quad \tilde{g}'_{ij}=\tilde{g}_{kl}\left(J^{-1}\right)^k_{~i}\left(J^{-1}\right)^l_{~j}.\nonumber
\eeq
\item{} $(v',u',x'^i)=(v+h(u,x^k),u,x^i)$
\beq
H'=H-h_{,u}, \quad W'_i=W_i-h_{,i}, \quad \tilde{g}_{ij}'=\tilde{g}_{ij}.\nonumber
\eeq
\item{} $(v,u,x^i)=(v/g_{,u}(u),g(u),x^i)$
\beq
H'=\frac 1{g_{,u}^2}\left(H+v\frac{g_{,uu}}{g_u}\right), \quad W'_i=\frac 1{g_{,u}}W_i, \quad \tilde{g}'_{ij}=\tilde{g}_{ij}.\nonumber
\eeq
\end{enumerate}

\subsection{ {$CSI$}$_0$ spacetimes}
The linearly independent components of the Riemann tensor with boost weight 1 and 0 are:
\beq
R_{\hat 0\hat 1\hat 0\hat i}&=&-\frac 12 W_{\hat i,vv}, \\
R_{\hat 0\hat 1\hat 0\hat 1}&=& -H_{,vv}+\frac 14\left(W_{\hat{i},v}\right)\left(W^{\hat i,v}\right), \\
R_{\hat 0\hat 1\hat i\hat j}&=& W_{[\hat i}W_{\hat j],vv}+W_{[\hat i;\hat j],v}, \\
R_{\hat 0\hat i\hat 1\hat j}&=& \frac 12\left[-W_{\hat j}W_{\hat i,vv}+W_{\hat i;\hat j,v}-\frac 12 \left(W_{\hat i,v}\right)\left(W_{\hat j,v}\right)\right], \\
R_{\hat i\hat j\hat n\hat m}&=&\tilde{R}_{\hat i\hat j\hat n\hat m}.
\eeq
Hence, the spacetime is a {$CSI$}$_0$ spacetime if there exists a frame $\left\{ \ell, {\bf n}, {\bf m}^{\hat i}\right\}$, a constant $\sigma$, anti-symmetric matrix ${\sf a}_{\hat i\hat j}$, and symmetric matrix ${\sf s}_{\hat i\hat j}$ such that
\beq
W_{\hat i,vv} &=& 0, \label{Wcsi1}\\
H_{,vv}-\frac 14\left(W_{\hat i,v}\right)\left(W^{\hat i,v}\right) &=& \sigma, \\
W_{[\hat i;\hat j],v} &=& {\sf a}_{\hat i\hat j}, \\
W_{(\hat i;\hat j),v}-\frac 12 \left(W_{\hat i,v}\right)\left(W_{\hat j,v}\right) &=& {\sf s}_{\hat i\hat j},
\label{Wcsi4}\eeq
and the components $\tilde{R}_{\hat i\hat j\hat m\hat n}$ are all constants (i.e., $\d S^2_H$ is curvature homogeneous).

We note that the first equation implies that
\beq
W_{\hat i}(v,u,x^k)=v{W}_{\hat i}^{(1)}(u,x^k)+{W}_{\hat i}^{(0)}(u,x^k),
\eeq
while the second implies
\beq
H(v,u,x^k)=\frac{v^2}{8}\left[4\sigma+({W}_i^{(1)})({W}^{(1)i})\right]+v{H}^{(1)}(u,x^k)+{H}^{(0)}(u,x^k).
\eeq
If $\d S^2_H=\delta_{\hat i\hat j}{\bf m}^{\hat i}{\bf m}^{\hat j}$ is a (locally) homogeneous
Riemannian space, then there exists a frame where  $\tilde{R}_{\hat i\hat j\hat m\hat n}$ are all
constants. In general, curvature homogeneous does not imply homogeneous; however, 
since we are mostly interested in the {$CSI$} case, in light of Theorem \ref{thm:Kundt},  
we will henceforth take $\d S^2_H$ as a (locally) homogeneous metric.

\subsection{ {$CSI$}$_1$ spacetimes}
For a {$CSI$}$_0$ spacetime, we have
\beq
R_{\hat 0\hat 1\hat 0\hat i;\hat 1}&=&-\frac 12\left[\sigma W_{\hat i,v}-\frac 12({\sf s}_{\hat j\hat i}+{\sf a}_{\hat j\hat i})W^{\hat j,v}\right], \\
R_{\hat 0\hat i\hat j\hat k;\hat 1}&=&-\frac 12\left[W^{\hat n,v}\tilde{R}_{\hat n\hat i\hat j\hat k}-W_{\hat i,v}{\sf a}_{\hat j\hat k}+({\sf s}_{\hat i[\hat j}+{\sf a}_{\hat i[\hat j})W_{\hat k],v}\right].
\eeq
For the spacetime to be {$CSI$}$_1$, it is sufficient to require that the above components are constants; i.e.
\beq
\sigma W_{\hat i,v}-\frac 12({\sf s}_{\hat j\hat i}+{\sf a}_{\hat j\hat i})W^{\hat j,v} &=& {\mbold\alpha}_{\hat i}, \\
\left({\sf s}_{\hat i\hat j}+{\sf a}_{\hat i\hat j}\right)_{;\hat k}-\left({\sf s}_{\hat i\hat k}+{\sf a}_{\hat i\hat k}\right)_{;\hat j}&=& {\mbold\beta}_{\hat i\hat j\hat k},
\eeq
where the Ricci identity has been used to rewrite the latter condition.

\section{Examples of {$CSI$} spacetimes of Kundt form}
Various classes of {$CSI$} spacetimes 
that arise as members of $\FC \bigcap \KC$
can now be found. The solutions come in classes according to the properties of $W^{(1)}_{\hat{i}}(u,x^k)$.  In the following, $\mathcal{M}_H$ is a locally homogeneous Riemannian space.
\subsection{$W^{(1)}_{\hat{i}}(u,x^k)=0$}
Assuming $W^{(1)}_{\hat{i}}(u,x^k)=0$ immediately makes the metric (\ref{Kundtmetric}) a {$CSI$} space.

A special subcase of this class which is worth mentioning is the Brinkmann metrics for which $H$ and $W_i$ are all independent of $v$; thus $\sigma={\sf s}_{ij}={\sf a}_{ij}=0$. In this case, the expressions for the curvature tensors simplify drastically. In particular, if $F_{\hat{j}\hat{i}}\equiv 2W_{[\hat{i};\hat{j}]}$, the Ricci tensor is
\beq
R_{\hat{1}\hat{1}}=\tilde{\Box}H-\frac 14\tilde{F}^2, \quad R_{\hat{1}\hat{i}}=\tilde{\nabla}^{\hat{j}}F_{\hat{j}\hat{i}}, \quad R_{\hat{i}\hat{j}}=\tilde{R}_{\hat{i}\hat{j}}.
\eeq

\subsection{$W^{(1)}_{\hat{i}}(u,x^k)=\mathrm{constant}$} This case is {$CSI$} if and only if the simpler metric, $\widetilde{\mathcal{M}}$:
\beq
\widetilde{\d s^2}=2\d u\left(\d v+\frac{v^2}{2}\tilde{\sigma}\d u+vW_{\hat i}^{(1)}{\bf m}^{\hat i}\right)+\delta_{\hat i\hat j}{\bf m}^{\hat i}{\bf m}^{\hat j},
\eeq
is a homogeneous space. The curvature invariants of the Kundt metrics in this class will have the same invariants as $\widetilde{\mathcal{M}}$; i.e. $\mathcal{I}=\mathcal{I}(\widetilde{\mathcal{M}})$.

The homogeneous space $\mathcal{M}_H$ can be considered as a quotient space  $\mathcal{M}_H=G/H$, 
where $G$ is the isometry group and $H$ is the isotropy group. The corresponding 
Lie algebra decomposition is $\mf{g}=\mf{h}+\mf{m}$, and we let this correspond to the algebra of left-invariant vectors. We note that if there exists a subalgebra $\mf{k}\subset\mf{m}$ such that $[\mf{k}^{\perp},\mf{k}]=0$, where $\mf{k}^{\perp}$ is the complement of $\mf{k}$ in $\mf{g}$, then we can choose an $\mathrm{Ad}(H)$-invariant ${\bf W}$:
\[ {\bf W}\in \mf{k}.\]
By letting  $W^{(1)}_{\hat{i}}{\bf m}^{\hat{i}}$ be its corresponding left-invariant one-form, the associated Kundt spacetime will be a {$CSI$}.

Note that some homogeneous spaces may be represented by inequivalent Lie algebra decompositions. 
For example, hyperbolic space, $\mb{H}^n$, can both be considered as the quotient $SO(1,n)/SO(n)$ 
and as a solvable Lie group $G$ \cite{Hervik}. In the latter case the decomposition gives $\mf{g}=\mf{m}$ and $\mf{h}=0$; hence, $[\mf{h},\mf{m}]=0$.

\subsection{$W^{(1)}_{\hat{i}}(u,x^k)=\phi_{,\hat{i}}=\mathrm{non-constant}$}
\subsubsection{$\mathcal{I}=\mathcal{I}(\mathrm{dS}\times \mathcal{M}_H)$} \label{subsect:dSM} $\sigma>0$:
\beq
\d s^2&=&\cos^2(\sqrt{\sigma}x)\left[2\d u\left(\d v+\widetilde{H}\d u+\widetilde{W}_i{\bf m}^i\right)\right]\nonumber \\  && +\d x^2+\frac{1}{\sigma}\sin^2(\sqrt{\sigma}x)\d S^2_{S^n} +\d S^2_{H},
\label{eq:csi2}\eeq
where
\beq
&& W^{(1)}_{\hat{x}}=2\sqrt{\sigma}\tan(\sqrt\sigma x),\nonumber \\
&& \widetilde{W}_i = \widetilde{W}_i^{(0)}(u,x^j), \quad
\widetilde{H}=\frac{v^2}{2}\sigma+v\widetilde{H}^{(1)}(u,x^j)+\widetilde{H}^{(0)}(u,x^j).
\eeq
The Riemann tensor is of type:
\[ R=(R)_0+(R)_{-1}+(R)_{-2}. \]
Special cases:
\begin{enumerate}
\item{} If $\widetilde{W}^{(0)}_i=\frac{1}{\sigma}\widetilde{H}^{(1)}_{~~,i}$ then $(R)_{-1}=0$.
\item{} If $\widetilde{W}^{(0)}_i=\widetilde{H}^{(1)}=\widetilde{H}^{(0)}=0$ 
then $(R)_{-1}=(R)_{-2}=0$ and $\mathrm{dS}\times \mathcal{M}_H$.
\end{enumerate}

\subsubsection{$\mathcal{I}=\mathcal{I}(\mathrm{AdS}\times \mathcal{M}_H)$}\label{subsect:AdSM}  $\sigma<0$:
\beq
\d s^2_I&=&\cosh^2(\sqrt{|\sigma|}x)\left[2\d u\left(\d v+\widetilde{H}\d u+\widetilde{W}_i{\bf m}^i\right)\right]\nonumber \\  && +\d x^2+\frac{1}{|\sigma|}\sinh^2(\sqrt{|\sigma|}x)\d S^2_{S^n} +\d S^2_{H},
\label{eq:csi3}\eeq
where
\beq
&&W^{(1)}_{\hat{x}}=-2\sqrt{|\sigma|}\tanh(\sqrt{|\sigma|} x), \nonumber \\
&&\widetilde{W}_i = \widetilde{W}_i^{(0)}(u,x^j), \quad
\widetilde{H}=-\frac{v^2}{2}|\sigma|+v\widetilde{H}^{(1)}(u,x^j)+\widetilde{H}^{(0)}(u,x^j).
\eeq
The Riemann tensor is of type:
\[ R=(R)_0+(R)_{-1}+(R)_{-2}. \]
Special cases:
\begin{enumerate}
\item{} If $\widetilde{W}^{(0)}_i=-\frac{1}{|\sigma|}\widetilde{H}^{(1)}_{~~,i}$ then $(R)_{-1}=0$.
\item{} If $\widetilde{W}^{(0)}_i=\widetilde{H}^{(1)}=\widetilde{H}^{(0)}=0$ 
then $(R)_{-1}=(R)_{-2}=0$ and $\mathrm{AdS}\times \mathcal{M}_H$.
\end{enumerate}

Let us present another example:
\beq
\d s^2_{II}&=&\sinh^2(\sqrt{|\sigma|}x)\left[2\d u\left(\d v+\widetilde{H}\d u+\widetilde{W}_i{\bf m}^i\right)\right]\nonumber \\  && +\d x^2+\frac{1}{|\sigma|}\cosh^2(\sqrt{|\sigma|}x)\d S^2_{\mathbb{H}^n} +\d S^2_{H},
\label{eq:csi4}\eeq
where
\beq
&&W^{(1)}_{\hat{x}}=2\sqrt{|\sigma|}\coth(\sqrt{|\sigma|} x), \nonumber \\
&&\widetilde{W}_i = \widetilde{W}_i^{(0)}(u,x^j), \quad
\widetilde{H}=\frac{v^2}{2}|\sigma|+v\widetilde{H}^{(1)}(u,x^j)+\widetilde{H}^{(0)}(u,x^j).
\eeq
The Riemann tensor is of type:
\[ R=(R)_0+(R)_{-1}+(R)_{-2}. \]
Special cases:
\begin{enumerate}
\item{} If $\widetilde{W}^{(0)}_i=\frac{1}{|\sigma|}\widetilde{H}^{(1)}_{~~,i}$ then $(R)_{-1}=0$.
\item{} If $\widetilde{W}^{(0)}_i=\widetilde{H}^{(1)}=\widetilde{H}^{(0)}=0$ then $(R)_{-1}=(R)_{-2}=0$.
\end{enumerate}

There are more {$CSI$} spacetimes with $\mathcal{I}=\mathcal{I}(\mathrm{(A)dS}\times \mathcal{M}_H)$. They are constructed similarly to the ones above. They are all related to the fact that de Sitter space or Anti-de Sitter space can be written in many ways as fibered spaces over the spheres, $S^n$, or the hyperbolic space, $\mathbb{H}^n$, respectively (see  Appendix \ref{app:4D} for the 4D case).

\subsection{$W^{(1)}_{\hat{a}}(u,x^k)=\mathrm{non-constant}, ~W^{(1)}_{\hat{\alpha}}(u,x^k)=\mathrm{constant}$}
This is a generalisation of the Warped {$CSI$} spacetimes considered earlier. The metric can be written
\beq
\d s^2&=&e^{-2\phi(x^\alpha)}\left[2\d u\left(\d v+\widetilde{H}\d u+\widetilde{W}_a\d x^a+\widetilde{W}_\alpha{\mbold{\omega}}^\alpha\right)+\delta_{ab}\d x^a\d x^b\right] \nonumber \\
&& +\delta_{\alpha\beta}{\mbold\omega}^\alpha{\mbold\omega}^\beta,
\label{eq:csi}\eeq
where
\beq
\widetilde{W}_a &=& v\widetilde{W}_a^{(1)}(u,x^a)+\widetilde{W}_a^{(0)}(u,x^a,x^\alpha), \\
\widetilde{W}_\alpha &=& \widetilde{W}_\alpha^{(0)}(u,x^a,x^\alpha), \\
\phi &=& \phi(x^\alpha), \quad \phi_{,\alpha}=\mathrm{constant}, \\
\widetilde{H}&=&\frac{v^2}{8}(\widetilde{W}^{(1)}_a)(\widetilde{W}^{(1)a})+v\widetilde{H}^{(1)}(u,x^a,x^\alpha)+\widetilde{H}^{(0)}(u,x^a,x^\alpha).
\eeq
Furthermore, we require that $\delta_{\alpha\beta}{\mbold\omega}^\alpha{\mbold\omega}^\beta=\tilde{g}_{\alpha\beta}(x^\sigma)\d x^\alpha \d x^\beta$ is a 
Riemannian homogeneous space, $\mathcal{M}_H$. The allowed forms of the function $\phi$ depend on the  homogeneous space $\mathcal{M}_H$. Topologically, this space is a fibered 
manifold with base manifold $\mathcal{M}_H$ and we will require that the fiber is a {$VSI$} space; 
in particular, this means that $\widetilde{W}_a$ satisfy the {$VSI$} equations 
(\ref{Wcsi1})-(\ref{Wcsi4}) with $\sigma={\sf a}_{ij}={\sf s}_{ij}=0$.
Special cases:
\begin{enumerate}
\item{} If $\widetilde{W}^{(0)}_{\hat{i}}=\Psi_{,\hat{i}}, ~ \widetilde{W}^{(0)}_i=\widetilde{H}^{(1)}=0$ then $(R)_{-1}=0$.
\item{} If $\widetilde{W}^{(0)}_{\hat{i}}=\widetilde{H}^{(1)}=\widetilde{H}^{(0)}=0$ then $(R)_{-1}=(R)_{-2}=0$.
\end{enumerate}

The connection $\nabla$ for the above metric can be decomposed as
\beq
\nabla=\widetilde{\nabla}-{\mbold{\tau}},
\eeq
where ${\mbold\tau}$ is a $(2,1)$ tensor (or operator acting in the obvious way) with only non-positive boost-weight components, while $\widetilde{\nabla}$ has the following property: {\sl There exists a null frame  such that the connection coefficients of $\widetilde{\nabla}$ has the following properties: }
\begin{enumerate}
\item{} Positive boost weight connection coefficients are all zero.
\item{} Zero boost weight connection coefficients are either all constants or are connection coefficients of $\mathcal{M}_H$.
\end{enumerate}
Henceforth we will assume that this null frame is chosen. Moreover, ${\mbold\tau}$ and $\widetilde{\nabla}$ can be chosen such that 
\beq
&&{\mbold\tau}_{{\bf e}_{\alpha}}X={\mbold\tau}_{X}{\bf e}_{\alpha}=0,\quad \nonumber \\
&& \widetilde{\nabla}_{X}g=0, \quad g({\mbold\tau}_XY,Z)+g(Y,{\mbold\tau}_XZ)=0, \quad \forall X,Y,Z\in TM.
\eeq
Note that the torsion of $\widetilde{\nabla}_{X}$ is $S_XY={\mbold\tau}_YX-{\mbold\tau}_XY$. 

Let $R$ and $\widetilde{R}$ be the curvature tensors to the connections $\nabla$ and $\widetilde{\nabla}$ respectively. Then
\beq
R=\widehat{R}+\widetilde{R},
\eeq
where $(\widehat{R})_b=0$ for $b\geq 0$  while $(\widetilde{R})_b=0$ for $b>0$. Moreover, all boost weight 0 components are constants\footnote{This can be seen by explicit calculation of the curvature tensors.}. We also have
\begin{thm}
For all $k$, $\nabla^{(k)}{R}$ has maximal boost order 0 where all boost weight 0 components are constants.
\end{thm}
\begin{proof}
We will prove the Theorem by induction. Let us in the following denote the projection of a tensor $T$ onto the vector space spanned by the components of boost weight $b$ by $(T)_b$. The key observation is that the $(R)_0=(\widetilde{R})_0$ is of constant curvature over the $VSI$ fibers. As a result, ${\mbold\tau}({R})_0=0$. Thus the theorem is true for ${\nabla}({R})_0$. Moreover, by applying the Bianchi identity for ${\nabla}{R}$, we can show that it is true for ${\nabla}({R})_{-1}$. Hence, the theorem is true for $k=1$.

Assume the theorem is true for $k$. Then consider $\nabla\nabla^{(k)}R$. Immediately, we have that it is true for $\widetilde{\nabla}({\nabla}^{(k)}{R})_0$. Moreover, the curvature tensor is of constant curvature over the $VSI$ fibers, so ${\mbold\tau}({\nabla}^{(k)}{R})_0=0$; hence, the theorem is true for ${\nabla}({\nabla}^{(k)}{R})_0$

For ${\nabla}({\nabla}^{(k)}{R})_{-1}$ the critical part is the one that raises the boost weight by one. However, by using the identity
\beq
\left[{\nabla}_{A},{\nabla}_{B}\right]T_{C_1\dots C_i\dots C_p}=\sum_{i=1}^p{R}^{E}_{~C_i AB}T_{C_1\cdots E\cdots C_p},
\label{commutator}\eeq
recursively, and also the Bianchi identity, we see that it is also true 
for ${\nabla}({\nabla}^{(k)}{R})_{-1}$. Hence, by induction, the Theorem follows.
\end{proof}
These results consequently imply that the 'fibered {\sc vsi}' 
spacetime given by eq.(\ref{eq:csi}) is a $CSI$ spacetime. A crucial part of the proof is 
that $(R)_0$ is of constant curvature over the $VSI$ fibers, which implies 
that ${\mbold\tau}({\nabla}^{(k)}{R})_0=0$. 

\begin{thm}
Consider the metric (\ref{eq:csi}). If for all $k$ $\nabla^{(k)}{R}$ has maximal 
boost order 0, where all boost weight 0 components are constants, then 
there exists a homoegenous space $(\widetilde{\mathcal{M}},\tilde{g})$ with the same 
curvature invariants as the metric (\ref{eq:csi}).
\end{thm}
\begin{proof}
The proof essentially follows along the same lines as above by noticing that the functions $\widetilde{W}_i$ and $\widetilde{H}$ do not contribute to the curvature invariants. The homogeneous space can be obtained by setting all of these to zero. 
\end{proof}

\section{Discussion}

We have obtained a number of general results.
We have studied warped product $CSI$ spacetimes (Theorem 3.1).
We have presented the canonical form for the Kundt $CSI$ metric
(Theorem 4.1), and studied higher dimensional  Kundt
spacetimes in some detail. In sections 3 \& 5 we have
constructed higher dimensional examples of
$CSI$ spacetimes that arise as warped products and that belong to the
Kundt class. In the appendices we shall present
all of the 3-dimensional $VSI$ metrics,
explicitly construct the metrics in
the 4-dimensional $CSI$ spacetimes, and we shall establish the
canonical higher dimensional $VSI$ metric.

Motivated by our results and examples we propose the following conjectures:

\begin{con}[$\FC$ conjecture] A spacetime is {$CSI$} iff there exists a null frame 
in which the Riemann tensor and its derivatives can be brought into one of the following forms: either
\begin{enumerate}
\item{} the Riemann tensor and its derivatives are constant, in which case we have a
locally homogeneous space, or
\item{} the Riemann tensor and its derivatives are of boost order zero with constant boost weight zero components at each
order.  This implies that Riemann tensor is of type II or less.
\end{enumerate}
\end{con}
Whereas the proof of the `if' direction is trivial, the `only if' part is significantly more difficult and would
require considering second order curvature invariants.  We also point out that
if we relax {$CSI$} to {$CSI$}$_{1}$ then in (1)
we must now include all 1-curvature homogeneous spacetimes as well, since these naturally
imply {$CSI$}$_{1}$ and are not
necessarily homogeneous spaces.

Assuming that the above conjecture is correct and that there exists such a preferred
null frame, then

\begin{con}[$\KC$ conjecture] If a spacetime is {$CSI$}, then
the spacetime is either locally homogeneous or belongs to the
higher dimensional Kundt $CSI$ class.
\end{con}

Finally, we have that
\begin{con}[$\RC$ conjecture]
If a spacetime is {$CSI$}, then it can be constructed from locally homogeneous
spaces and $VSI$ spacetimes.
\end{con}
This construction can be done by means of fibering, warping and tensor sums.

From the results above and these conjectures, it is plausible that for $CSI$
spacetimes that are not locally homogeneous, the Weyl type is
$II$, $III$, $N$ or $O$, and that all boost weight zero terms are
constant.

We intend to study the general validity of these conjectures in future work. The relationship
to curvature homogeneous spacetimes is addressed in \cite{Mp}.
Support for these conjectures in the 4D case is discussed in the
next section.

\section{Four--dimensional $CSI$}

Let us now consider the 4D $CSI$ spacetimes in more detail. We
assume that a $CSI$ spacetimes is {\em either} of Petrov
(P)-type I or Plebanski-Petrov (PP)-type I, {\em or} of P-type II
and PP-type II (branch A or B of Table 1,
respectively). In branch A, we explicitly assume that the
spacetime is not of P-type II and PP-type II, otherwise we would
be in branch B. It is plausible that in branch A the
spacetimes must be of P-type I \emph{and} PP-type I (but we have not
established this here). Since the spacetime is not of P-type II
and PP-type II, it is necessarily completely backsolvable (CB)
\cite{Carminati}. Since all of the zeroth order scalar curvature
invariants are constant, there exists a frame in which the
components of the Riemann tensor are all constant, and the
spacetime is curvature homogeneous $CH_0$ (i.e., $CSI_0 \equiv
CH_0$). If the
spacetime is also $CH_1$, then it is necessarily locally
homogeneous $H$ \cite{Mp}. Therefore, if it is not locally
homogeneous, it cannot be $CH_1$. Finally, by considering
differential scalar invariants, we can obtain information on the
spin coefficents in the $CSI$ spacetime. Although a comprehensive
analysis of the differential invariants is necessary, a
preliminary investigation indicates that $\tau, \rho$ and $\sigma$
must all be constant and are likely zero. Therefore, in branch A,
there are severe constraints on those spacetimes that are not
locally homogeneous, and it is plausible that there are no such
spacetimes.

In branch B of Table 1, we have that the spacetime is necessarily
of P-type II and PP-type II. It then follows from Theorem 7.1 below,
that all boost weight zero terms are necessarily constant
(i.e., the spacetime is $\FCSI{0}$). The spacetime is then
either CB, in which case all of the results in branch A apply, or
they are NCB and a number of further conditions apply (these
conditions are very severe \cite{Carminati}). In either case,
there are a number of different classes characterized by their
P-type and PP-type (in each case at least of type II), and in each
class there are a number of further restrictions. By investigating
the differential scalar invariants we then find conditions on the
spin coefficients. By considering each class separately, we can
then establish that $\tau = \rho = \sigma =0$, and that the
spacetime is necessarily $ \KC$. All of these
spacetimes are constructed in Appendix B.

\begin{table}
  \includegraphics[width=13cm]{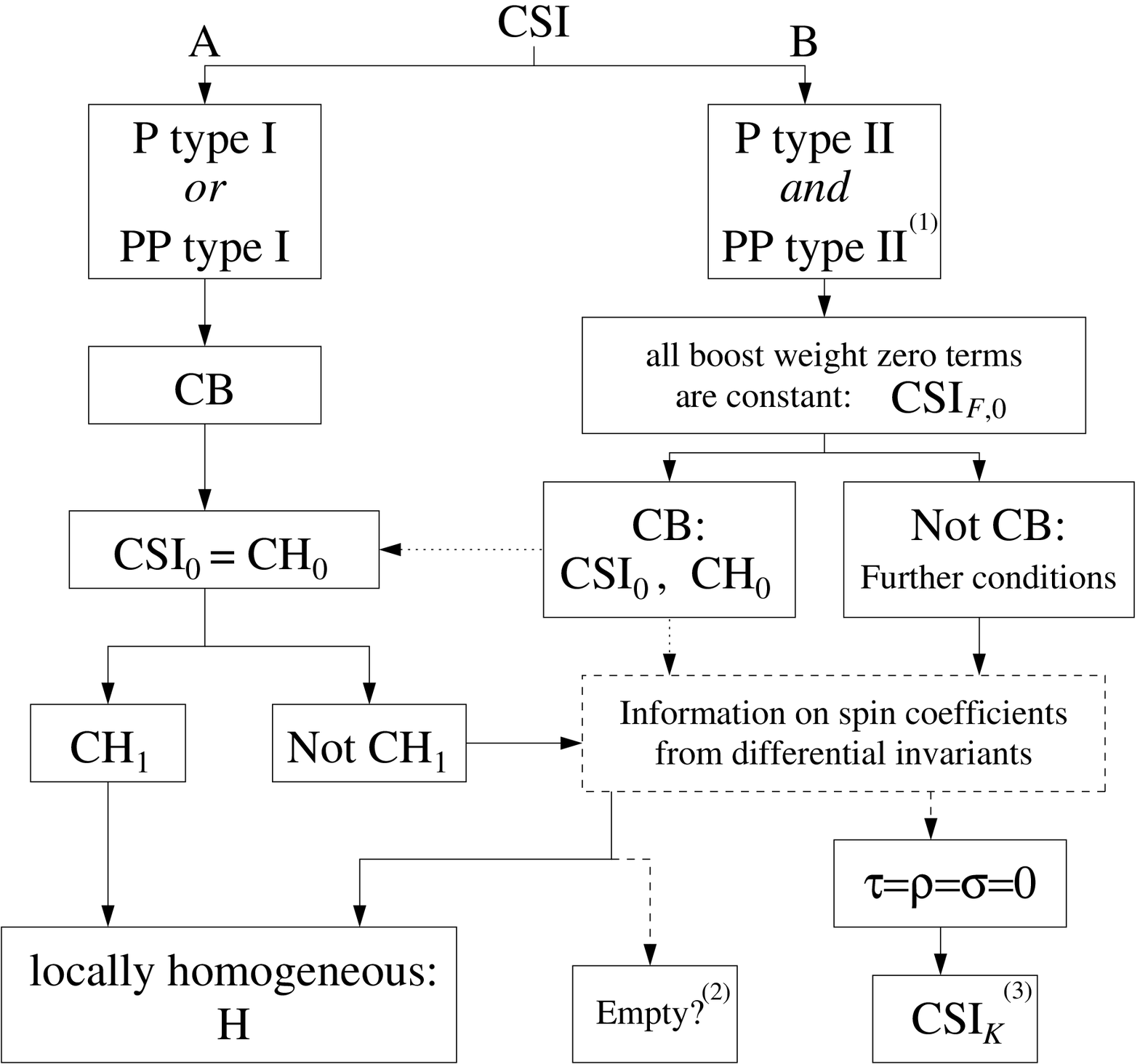}
  \caption{4D $CSI$. All terms are defined in the text. 
  A dashed line indicates a conjectured result. Footnotes:
(1) this indicates that the spacetime is at least of type $II$
(i.e., of type $II$, $III$, $N$ or $0$), (2) spacetimes in this set are
described in the text, (3)
all of the 4D $\KC$ spacetimes are displayed
explicitly in the text.}\label{Table}
\end{table}

The results are summarized in Table 1. In branch A of Table 1, we
have established that if the $CSI$ spacetime is not locally
homogeneous, then necessarily it is of P-type I or PP-type I (and
does {\em not} belong to $\KC$, $\FC$, or
$\RC$), it belongs to $CSI_0 \equiv CH_0$, but not
$CH_1$ and there are a number of further constraints arising from
the non-CB conditions and the differential constraints (i.e.,
constraints on the spin coefficients). This exceptional set is
very sparse, and mostly likely empty; but in order to demonstrate
this the conjectures in 4D must be proven, utilising a number of
4D invariants.

We have explicitly constructed the 4D $\FC$,
$\KC$, and $\RC$, and established the
relationship between these $CSI$ subsets themselves and with $CSI
\backslash H$, thereby lending support to the conjectures in the
previous section. The metrics of these $CSI$ spacetimes are
described in Appendix B. From this calculation of all of the
members of the set $\FC \bigcap \KC$, we see
that they all belong to $\RC$. That is, we have
established the uniqueness of $\RC$ within this
particular context. In particular, we have found all $CSI$ in
branch B of Table 1.

Finally, let us  consider Petrov (P)-type II and Plebanski-Petrov (PP)-type II
Lorentzian $CSI$ spacetimes. We shall show that if all zeroth order
curvature invariants are constant then the boost
weight zero curvature scalars, $\Psi_{2}$, $\Phi_{11}$ and $\Phi_{02}$ are
constant.  For these algebraically special
spacetimes the converse follows straightforwardly.

\begin{thm}
If all zeroth order
curvature invariants are constant in Petrov-type II and Plebanski-Petrov-type II
Lorentzian $CSI$ spacetimes, then the boost
weight zero curvature scalars, $\Psi_{2}$, $\Phi_{11}$ and $\Phi_{02}$ are
constant.
\end{thm}
\begin{proof}

We make use of the following curvature invariants
\begin{equation}
\begin{array}{ccc}
w_{1}=\Psi_{ABCD}\Psi^{ABCD}, &
r_{1}=\Phi_{AB\dot{A}\dot{B}}\Phi^{AB\dot{A}\dot{B}}, &
r_{2}=\Phi_{AB\dot{A}\dot{B}}\Phi^{B\quad\dot{B}}_{\quad
C\quad\dot{C}}\Phi^{CA\dot{C}\dot{A}}.
\end{array}
\end{equation}
It is worth noting that in general there is an additional independent
degree 4 Ricci invariant $r_{3}$; however, in
PP-types II or less there always exists a syzygy for $r_{3}$ in terms of
$r_{1}$ and $r_{2}$ (or else it vanishes).  This
is analogous to the well known syzygy for the Weyl invariants,
$I^{3}=27J^{2}$, signifying an algebraically special
spacetime.

Assuming first that the Weyl and Ricci canonical frames are aligned, then
the non-vanishing curvature scalars are
$\Psi_{2}$, $\Psi_{4}=1$, $\Phi_{11}$, $\Phi_{20}=\Phi_{02}$,
$\Phi_{22}=1$ and we have $w_{1}=6\Psi_{2}^2$,
$r_{1}=2\Phi_{02}^2+4\Phi_{11}^2$, $r_{2}=-6\Phi_{02}^2\Phi_{11}$. Clearly
$w_{1}$ constant implies $\Psi_{2}$
constant.  If $\Phi_{02}(2\Phi_{11}-\Phi_{02})(2\Phi_{11}+\Phi_{02})\neq0$
then $\Phi_{02}$ and $\Phi_{11}$ can be
expressed in terms of $r_{1}$ and $r_{2}$; therefore, constant $r_{1}$ and
$r_{2}$ gives constant $\Phi_{11}$ and
$\Phi_{02}$.  If $\Phi_{02}=0$ or $\Phi_{02}=\pm 2\Phi_{11}$ then $r_{1}$
can be used to give the same result.  In this
aligned case we find that constant $R$, $w_{1}$, $r_{1}$ and $r_{2}$
implies that all curvature scalars are constant
(including the boost weight 0 scalars); therefore, we have a curvature
homogeneous spacetime.

In the non-aligned case we refer to Carminati and Zakhary \cite{Carminati} where it was shown that for
this PP-type, complete backsolving (CB) of the
Carminati-Zakhary (CZ) invariants can be achieved and all curvature scalars
can be expressed in terms
of zeroth order invariants; thus constant CZ
invariants implies constant curvature scalars.  In the exceptional case in which
$\Psi_{0}=\Psi_{1}=0$ in the Ricci canonical frame,
complete backsolvability is not possible since $\Psi_{3}$ and $\Psi_{4}$
cannot be determined from any zeroth order
curvature invariant \cite{Carminati}.  However, this exceptional case occurs
when the Weyl and Ricci canonical frames are
aligned, for P-type II $\Psi_{3}=0$ and $\Psi_{4}=1$ and the remaining
curvature scalars were shown above to be
constant (this is a particular instance of a not completely backsolvable (NCB)
case becoming completely backsolvable; see
comment in Ref. 3 of \cite{Carminati}).
\end{proof}

We have consequently also shown that for P-type II and PP-type II spacetimes, if the zeroth order
invariants are constant then all curvature scalars
are constant. These results support the higher dimensional conjectures discussed
in the previous section.

\section*{Acknowledgements} We would like to thank R. Milson, V. Pravda and A. Pravdova 
for helpful comments. This work was supported by NSERC (AC and NP) and the Killam
Foundation (SH).
\newpage

\appendix

\section{3-dimensional $VSI$ metrics}

By analogy with the four dimensional $VSI$ spacetimes, we begin by
considering a real null frame $e_{i}=\{l,n,m\}$ such that the only
non-vanishing inner products are $l^{a}n_{a}=1=-m^{a}m_{a}$.  In
three dimensions, the Riemann tensor is equivalent to the Ricci
tensor, therefore all zeroth order invariants vanish if

\begin{equation}
R_{ab}=R_{22}l_{a}l_{b}+2R_{23}l_{(a}m_{b)}.
\end{equation}

We identify $R_{22} \sim \phi_{22}$ and $R_{23} \sim \phi_{12} +
\phi_{21}$ and we use the n-dimensional version of the $VSI$
theorem to conclude that the analogues of $\kappa$, $\sigma$ and
$\rho$ must vanish.  In three dimensions we find that that $\kappa
\sim \gamma_{311}$, $\sigma=\rho\sim \gamma_{313}$, and $\tau \sim
\gamma_{312}$, where $\gamma_{ijk}$ are the Ricci-rotation
coefficients.  Therefore, we have that a three dimensional
spacetime is $VSI$ if and only if $\gamma_{311}=\gamma_{313}=0$
and has Segre type \{3\}, \{(21)\} or \{(111)\} with the
possibility that the (non-)vanishing of $\gamma_{312}$ may lead to
distinct subclasses.

The method we use to explicitly obtain the three dimensional $VSI$
spacetimes involves writing the four dimensional $VSI$ spacetimes
in terms of real coordinates $(u,v,x,y)$ then considering the
resulting frame when either $x=const.$ or $y=const.$  We set
$\zeta=(x+i y)/\sqrt{2}$ and $W=(W_1 + i W_2)/\sqrt{2}$
throughout, and in every case the restriction to $y=const.$ yields
nothing new, hence the relevant null frame, in coordinates
$(u,v,x)$, is

\begin{eqnarray}
l=\partial_{v}, & n=\partial_{u}-[H+\frac{1}{2}(W_{1}^{2}+W_{2}^{2})]\partial_{v}+W_{1}\partial_{x}, & m=\partial_{x}
\end{eqnarray}

Upon restriction to $x=const.$ most of the four dimensional $VSI$
spacetimes become either flat or a special case of a null frame
given in the following tables. In addition, whenever
$\gamma_{312}$ is nonzero then it is always equal to $-1/x$.

\begin{table}
\begin{center}
\begin{tabular}{cl}
$\gamma_{312}$ & Constraints on $W_{1}(u,v,x)$, $W_{2}(u,v,x)$ and $H(u,v,x)$ \\ \hline \\

$=0$ & A1) (P-III ; $\tau = 0$ ; PP-N)\\
 & $W_{1}=W_{01}(u,x)$, $W_{2}=W_{02}(u,x)$, $H=vh_{1}(u,x)+h_{0}(u,x)$  \\
\\
$\neq 0$ & B1) (P-III ; $\tau \neq 0$ ; PP-N)\\
 & $W_{1}=-\frac{2v}{x}+W_{01}(u,x)$, $W_{2}=W_{02}(u,x)$, $H=-\frac{v^2}{2x^2}+vh_{1}(u,x)+h_{0}(u,x)$

\end{tabular}
\end{center}
\caption{Segre type \{3\} i.e. $R_{22} \neq 0$ and $R_{23} \neq 0$.}
\end{table}

\begin{table}
\begin{center}
\begin{tabular}{cl}
$\gamma_{312}$ & Constraints on $W_{1}(u,v,x)$, $W_{2}(u,v,x)$ and $H(u,v,x)$ \\ \hline \\

$=0$ & D1) (P-N ; $\tau = 0$ ; PP-O)  special case of A1) \\
 & $W_{1}=xf_{1}(u)+f_{0}(u)$, $W_{2}=xg_{1}(u)+g_{0}(u)$, $H=h_{0}(u,x)$  \\
\\
$\neq 0$ & F1) (P-III ; $\tau \neq 0$ ; PP-O) special case of B1) \\
 & $W_{1}=-\frac{2v}{x}+W_{01}(u)$, $W_{2}=W_{02}(u)$, $H=-\frac{v^2}{2x^2}+\frac{v}{x}W_{01}(u)+h_{0}(u,x)$

\end{tabular}
\end{center}
\caption{Segre type \{(21)\} i.e. $R_{22} \neq 0$ and $R_{23} = 0$.}
\end{table}

\begin{table}
\begin{center}
\begin{tabular}{cl}
$\gamma_{312}$ & Constraints on $W_{1}(u,v,x)$, $W_{2}(u,v,x)$ and $H(u,v,x)$ \\ \hline \\

$=0$ & I1) (P-III ; $\tau = 0$ ; vacuum)\\
 & $W_{1}=W_{01}(u)$, $W_{2}=W_{02}(u)$, $H=xh_{01}(u)+h_{02}(u)$  \\
\\
$\neq 0$ & L1) (P-N ; $\tau \neq 0$ ; vacuum)\\
 & $W_{1}=-\frac{2v}{x}$, $W_{2}=0$, $H=-\frac{v^2}{2x^2}+\sqrt{2}x[xh_{01}(u)+h_{02}(u)]$

\end{tabular}
\end{center}
\caption{Segre type \{(111)\} i.e. $R_{ij}=0$.}
\end{table}

Tables 2--4 provide a list of the three-dimensional $VSI$ spacetimes. 
Here, we have chosen to characterise the 3D cases according to Segre type. In this regard, the classification in \cite{HallCapocci} of three dimensional spacetimes is worth noting. In 4D the Segre types are related to the PP types \cite{exsol}. We have also indicated the corresponding class of four dimensional $VSI$
spacetimes (in brackets) from which the 3D solutions were obtained. It is also worth noting that
some of the spacetimes given here may be equivalent and so these
tables may reduce further. For example, in Table 4 the vanishing
of the Ricci tensor implies that all spacetimes listed there are
flat; therefore, using the full 3 parameter Lorentz group,
$SO(1,2)$, it should be possible to set $\gamma_{312}=0$ (i.e.,
the flat metric should appear in this table). In Tables 2 and 3
it is necessary to consider the frame freedom left after the Ricci tensor has
been brought to its canonical form. This can then be used to
determine if $\gamma_{312}$ can be made to vanish; alternatively,
it may be an invariant of this left over frame freedom and the
metrics would be inequivalent. If $\gamma_{312}$ can be set to
zero, then it still remains to be shown if a coordinate
transformation can be found relating the two metrics.


\section{4D $CSI$}
\label{app:4D}

We know that (Lorentzian) homogeneous spaces and $VSI$ spacetimes are
necessarily $CSI$ spacetimes. Let us display all 4D spacetimes
that are $\FC$ and $\KC$. We shall find
that all such spacetimes are necessarily in $\RC$.

The 4D spacetimes in $\FC  \bigcap \KC$ are
as follows.

\subsection{}\label{AppB1} $\mathcal{I}(\mathrm{AdS}_4)$
\beq
\d s^2&=&e^{-2py}\left[2\d u\left(\d v+\widetilde{H}\d u+\widetilde{W}_x\d x+\widetilde{W}_y\d y\right)+\d x^2\right] +\d y^2,
\label{eq:4Dcsi}\eeq
where
\beq
\widetilde{W}_x &=& v\widetilde{W}_x^{(1)}(u,x)+ \widetilde{W}_x^{(0)}(u,x,y),\nonumber  \\
\widetilde{W}_y &=& \widetilde{W}_y^{(0)}(u,x,y),\nonumber \\
\widetilde{H}&=&\frac{v^2}{8}(\widetilde{W}^{(1)}_x)^2+v\widetilde{H}^{(1)}(u,x,y)+\widetilde{H}^{(0)}(u,x,y).\nonumber
\eeq and $\widetilde{W}^{(1)}_x$ satisfy the 3D {$VSI$}
equations (see Appendix A).

Since $\widetilde{W}_x^{(0)}(u,x,y)$ depends on $y$, in addition
to $u$ and $x$, this metric is fibered.  Due to the terms
$e^{-2py}$ and $\d y^2$ in the metric, the spacetime is a warped product. By
redefining $\widetilde{W}_x$ in the metric above by omitting the
term $v\widetilde{W}_x^{(1)}(u,x)$ (the '$vW$' term) and in
$\widetilde{H}$ omitting the term
$\frac{v^2}{8}(\widetilde{W}^{(1)}_x)^2$ (the '$v^2H$' term), we
can rewrite the metric as the tensor sum of the redefined metric
(\ref{eq:4Dcsi}) and the metric
\beq \d s^2{_+}&
=&e^{-2py}\left[2\d u\left(\d v+
\frac{v^2}{8}(\widetilde{W}^{(1)}_x)^2 \d u+
v\widetilde{W}_x^{(1)}\d x \right)+\d x^2\right] +\d y^2,
\label{eq:4Dcsisum} \eeq 
Clearly the metric can be written in
terms of the combined operations of fibering, warping and tensor summing,
and hence belongs to $\RC$.

\subsection{} $\mathcal{I}({\sf Sol})$\beq
\d s^2&=&e^{-2py}\left[2\d u\left(\d v+\widetilde{H}\d u+\widetilde{W}_x\d x+\widetilde{W}_y\d y\right)\right]+e^{-2qy}\d x^2 +\d y^2,
\label{eq:4Dcsi3}\eeq
where
\beq
\widetilde{W}_x &=& \widetilde{W}_x^{(0)}(u,x,y),\quad
\widetilde{W}_y = \widetilde{W}_y^{(0)}(u,x,y),\nonumber \\
\widetilde{H}&=&v\widetilde{H}^{(1)}(u,x,y)+\widetilde{H}^{(0)}(u,x,y).\nonumber
\eeq This metric is fibered and warped.

\subsection{} $\mathcal{I}(M_1\times M_2)$:
\beq
\d s^2&=&2\d u\left(\d v+{H}\d u+{W}_i{\bf m}^i\right)+\delta_{ij}{\bf m}^i{\bf m}^j,
\label{eq:4Dcsi4}\eeq
where
\beq
{W}_i  &=& vW^{(1)}_i+\widetilde{W}_i^{(0)}(u,x,y),\nonumber \\
H &= &
v^2H^{(2)}+v\widetilde{H}^{(1)}(u,x,y)+\widetilde{H}^{(0)}(u,x,y).\nonumber
\eeq and  $W^{(1)}_i$ and $H^{(2)}$ are given in the table below.
This metric is fibered and warped, and due to the '$vW$' and
'$v^2H$' terms, can be written as a tensor sum.

\begin{tabular}{|c||c|c|c|}\hline
$\delta_{ij}{\bf m}^i{\bf m}^j$ & $\d x^2+\d y^2$ & $a^2(\d x^2+\sin^{2}x\d y^2)$ & $\d x^2+e^{-2qx}\d y^2$ \\
\hline \hline
$W^{(1)}_i{\bf m}^i$ & $\alpha \d x+\beta \d y$ & $0$& $\alpha \d x +\beta e^{-qx}\d y$\\
$H^{(2)}$ & $\frac 18(4\sigma+\alpha^2+\beta^2)$ &$\frac \sigma{2}$ & $\frac 18(4\sigma+\alpha^2+\beta^2)$ \\
\hline
\end{tabular}

\subsection{} $\mathcal{I}(\mathrm{(A)dS}_3\times \mathbb{R})$:
Metrics of the form (\ref{eq:csi2}), (\ref{eq:csi3}) and
(\ref{eq:csi4}). These metrics are fibered, warped and tensor
summed.

\subsection{}\label{AppB5}$\mathcal{I}(\mathrm{(A)dS}_4)$: Metrics of the form (\ref{eq:csi2}), (\ref{eq:csi3}) 
and (\ref{eq:csi4}). In addition, for  $\mathcal{I}(\mathrm{dS}_4)$
\beq \d
s^2&=&\cos^2(\sqrt{\sigma}y)\cos^2(\sqrt{\sigma}x)\left[2\d
u\left(\d v+\widetilde{H}\d u+\widetilde{W}_i{\bf
m}^i\right)\right]\nonumber \\ && +\cos^2(\sqrt{\sigma}y)\d x^2+\d
y, \eeq where \beq \widetilde{W}_i = \widetilde{W}_i^{(0)}(u,x,y),
\quad
\widetilde{H}=\frac{v^2}{2}\sigma+v\widetilde{H}^{(1)}(u,x,y)+\widetilde{H}^{(0)}(u,x,y).
\eeq and in the case of $\mathcal{I}(\mathrm{AdS}_4)$, similar
versions of the metrics (\ref{eq:csi3}) and (\ref{eq:csi4}) are obtained. These
metrics are fibered, warped and tensor summed.

All of the metrics in sections \ref{AppB1}-\ref{AppB5} can be written in terms of the
combined operations of fibering, warping and tensor summing.
Therefore, all of the spacetimes constructed belong to
$\RC$. We consequently have the result that in 4D
$\FC$, $\KC$ and $\RC$ are
equivalent. It is plausible that a similar result applies in
higher dimensions.

Finally, let us consider the tensor sum in a little more detail.
Let us consider the $\RC$ metric (\ref{eq:csi2}).
Redefining the metric by omitting the '$v^2H$' term, it can be
rewritten as the 'tensor sum' of the redefined metric and
\beq 
\d s^2{_+}&=& \left[2\d u\left(\d
v+\frac{v^2}{2}\sigma \d u + 2\sqrt{\sigma}\tan(\sqrt\sigma x) \d
x\right)\right] +\d x^2. \label{eq:csi2sum}\eeq The term
$\left[2\d u\left(\d v+\frac{v^2}{2}\sigma \d u\right)\right]$ can
be rewritten as $\left[\left(1+ \frac{\sigma}{2}UV\right)^{-2} \d U \d
V\right]$, which is a metric of constant Ricci scalar curvature.

It is plausible that we have constructed all of the 4D $CSI$ spacetimes here.
To prove this, we must prove the conjectures outlined above. This is perhaps best
done by considering higher order differential scalar
invariants in 4D \cite{4DVSI}, which we hope to do in a future
paper.

\section{VSI metrics}
We can now  write down the metric
for higher dimensional VSI spacetimes in a canonical form. An immediate corollary of Theorem \ref{thm:Kundt} is:
\begin{cor}
Any {$VSI$} metric can be written
\beq
\d s^2=2\d u\left[\d
v+H(v,u,x^k)\d u+W_{i}(v,u,x^k)\d x^i\right]+\delta_{ij}\d x^i\d x^j.
\eeq
\end{cor}
\begin{proof}
In \cite{Higher} it was proven that all {$VSI$} metrics have $L_{ij}=0$ and $R_{\hat{i}\hat{j}\hat{k}\hat{l}}=0$. The first condition implies that the metric is Kundt, hence of the form (\ref{Kundt}), while the second implies that the spatial metric, $\tilde{g}_{ij}$, is flat. Using Theorem \ref{thm:Kundt} the corollary now follows.
\end{proof}

We are now in a position to determine the explicit metric forms
for higher dimensional $VSI$ spacetimes \cite{CMPPPZ}, which we
shall return to in future work.

\end{document}